# Interfacial friction-induced electronic excitation mechanism for tribo-tunneling current generation

*Jun Liu\*, Keren Jiang, Lan Nguyen, Zhi Li, and Thomas Thundat\**

Dr. Jun Liu, Prof. Thomas Thundat
Department of Chemical and Biological Engineering, University at Buffalo, The State University of New York, New York 14260-4200, USA
E-mail: jliu238@buffalo.edu (Dr. Jun Liu); tgthunda@buffalo.edu (Prof. Thomas Thundat)
Dr. Keren Jiang, Lan Nguyen, and Dr. Zhi Li
Department of Chemical and Materials Engineering, University of Alberta, Edmonton, Alberta T6G2 V4, Canada

**Abstract**

Direct-current (d.c.) electricity generation using moving Schottky contact is emerging as a new strategy for mechanical energy conversion. Here, we demonstrate the generation of d.c. tunneling current with a density of ~35 A/m$^2$ at a metal-insulator-semiconductor (MIS) sliding system using micro-tips. The measured current densities were found to be three to four orders of magnitude higher than that observed with the conventional polymer-based triboelectric nanogenerators (TENGs). The electromotive force $\Delta V_s^*$ for the tribo-tunneling transport comes from the dynamic electronic excitation at the frictional interface rather than from the electrostatically trapped surface charges as in the case of conventional TENGs. The strong electronic excitation can give rise to a non-equilibrium interfacial charge variation $\Delta\sigma_m$. Depending on the energy distribution of the excited electrons/holes, $\Delta\sigma_m$ subsequently dissipates non-adiabatically into tunneling current and trapped surface charges, or adiabatically into heat. These fundamental results not only enhance our understanding of the triboelectric phenomenon, but also open up new paths for the development of next-generation mechanical energy harvesting and sensing techniques.

**Main Text**

Understanding the carrier dynamics at a frictional interface is essential for developing novel mechanical energy conversion technologies.[1-3] To date, d.c. generation phenomenon has been observed in various metal-semiconductor moving systems (metal-MoS$_2$ layers[4], metal-Si[5], graphene-Si[6], metal-conducting polymer (PPy)[7], SnO$_2$-PPy[8], *etc.*), and different mechanisms have been proposed to explain the generation of high density currents. Unlike a conventional triboelectric nanogenerator (TENG) that generates alternating current by dielectric displacement (a.c., $J$~0.01-0.1 A/m$^2$), sliding metal-semiconductor frictional contact induces strong localized electric field and produces high current density (d.c., $J$ of the order of 10 A/m$^2$) through quantum mechanical tunneling.[4, 5] Particularly, it has been found that the electromotive force (*emf*, measured in volts) in a tribo-tunneling system is related to an electronic excitation process under non-equilibrium condition, which cannot be explained by the electrostatics-based TENG theory.[4, 5] Though the triboelectric charge transfer phenomenon via contact electrification has been investigated for years, little is known for the electronic excitation due to frictional contact.[1, 9]

Generally, the open-circuit voltage ($V_{oc}$) in a conventional TENG is attributed to the separation of two dielectric materials carrying opposite electrostatic charges.[3, 10] As illustrated in **Figure 1**a, when the two bodies in contact are separated horizontally, the induced image charges on the electrodes, positive on one electrode and negative on another, provide the *emf* for the current flow in the external circuit.[4, 5] In contrast, the rising interfacial potential $\Delta V_s^*$ in the MS sliding contact (**Figure 1**b) shows very different features: i) $\Delta V_s^*$ is excited at an intimately contacted M-S interface in the absence of changing contact area or separation distance (capacitance $C$ remains constant), and ii) the sustained d.c. output arises instantaneously irrespective of the sliding direction.[4, 5] However, the $V_{oc}$ is negligible when the two contact bodies are intimately contacted in conventional TENGs (**Figure 1**a): the positive/negative electrostatic charges are confined in a 'two-dimensional' interface cancelling each other, and therefore electrodes are in isopotential.[3]

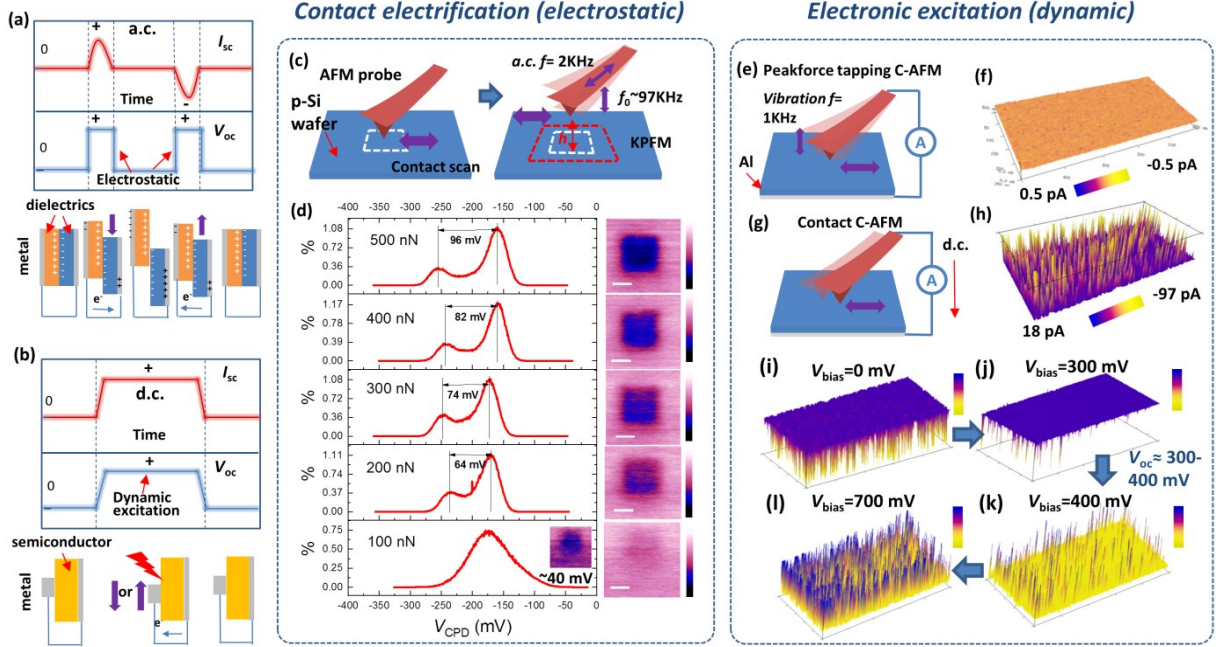

**Figure 1.** (a) Schematics of the short-circuit current $I_{sc}$ and the open-circuit voltage output $V_{oc}$ of conventional polymer-based TENGs. (b) Schematics of the $I_{sc}$ and the $V_{oc}$ of MS-based trbo-tunneling generator. (c) Schematics of the KPFM study of triboelectric charge accumulation. Contact mode scan is followed by Peakforce-KPFM characterization. The rubbed area is marked by dashed square. Peakforce setpoint: 5 nN; Scan rate: 1 Hz; Boron-doped diamond conductive AFM probe resonance frequency $f_0$=97 kHz, spring constant $k$= 5.4 N/m. (d) Contact force $F$-dependent KPFM surface potential distribution. (e) and (g) illustrate the principle of Peakforce tapping C-AFM and contact-C-AFM mode, respectively. (f) and (h) show the tunneling current signal under Peakforce tapping mode (Peakforce=300 nN) and contact mode ($F$=300 nN), respectively. The tribo-tunneling current is only generated under continuous sliding. (i-k) External bias-dependent C-AFM current output with $F$=500 nN; The $V_{oc}$ is determined to be 300~400 mV.

For comparison, we first used Kelvin probe force microscopy (KPFM) to investigate the triboelectric charging effect in conventional TENGs.[11, 12] In the AFM experiments, a p-type Si substrate (boron-doped, orientation <100>, resistivity: 0.1-1 Ω·cm) coated with aluminum bottom electrode was used. First, a selected area was rubbed with a boron-doped diamond AFM tip, followed by a Peakforce tapping-KPFM scan on an enlarged area (**Figure 1**c).[12, 13, 14] In KPFM measurement, $V_{CPD}$ is expressed by[13, 15]:

$$V_{\text{CPD}} = V_{\text{sample}} - V_{\text{tip}} = \frac{\varphi_{\text{tip}} - \varphi_{\text{sample}}}{q}, \quad (1)$$

where $V_{\text{sample}}$, $V_{\text{tip}}$, $\varphi_{\text{tip}}$, and $\varphi_{\text{sample}}$ denotes sample surface potential, tip surface potential, tip work function, and sample work function, respectively. Specially, the work function of a non-metallic material surface may be expressed in the form of[16]:

$$\varphi_{\text{sample}} = (E_c - E_F)_{\text{bulk}} - eV_s + \chi - \Delta\phi_s, \quad (2)$$

where $E_c$, $E_F$, and $(E_c - E_F)_{\text{bulk}}$ represent conduction band, Fermi energy level, and their difference in the bulk, respectively. $V_s$ is the surface band bending due the charged surface states, $\chi$ is the electron affinity, and $\Delta\phi_s$ is the short-range surface electrical dipole (see **Figure S1** in Supporting Information for details).

As shown in **Figure 1**d, the contrast of contact potential difference, $V_{CPD}$, can be observed on the rubbed area, which is associated with the increased local surface charge density ($\Delta\sigma$). Considering the electrical routing in the experiment (from sample to tip), a larger $\varphi_{\text{sample}}$ and hence a smaller $V_s$ is predicted on the rubbed area, indicating positive charge (*i.e.* hole) accumulation at the native silicon oxide surface. As the contact force $F$ increases, it can be seen that $\Delta V_{CPD}$ of the rubbed area increases from ~40 mV to ~96 mV, which can be considered as due to the increased microscopic contact area between the tip and the substrate, and the enhanced charge transfer under compression between the tip and the substrate surface atoms (**Figure 1**d).[17, 18] In dielectric materials based TENGs, those trapped surface charges induce image charges on the bottom electrode, generating voltage when the contact materials are separated vertically in distance or horizontally in area.

We also investigated the characteristics of $\Delta V_s^*$, which accounts for the tribo-tunneling current. **Figure 1**f and **1**h show the C-AFM tunneling current signal under Peakforce tapping mode (**Figure 1**e) and contact mode (**Figure 1**g), respectively. Under the Peakforce tapping mode, the AFM probe and sample are intermittently brought into contact while the tip is scanned across the sample off resonance at 1 kHz (the resonance frequency is ~97 kHz), and with the feedback loop controlling the maximum force on the tip (Peak force) for each individual cycle (See **Figure S**4 in the Supporting Information). It can be seen that C-AFM current is generated only under continuous sliding of the contact (**Figure 1**h), indicating that a strong lateral interfacial atomic interaction is required for triggering the electronic excitation. It should be noted that the negative current direction here refers to the current flowing from the tip to the sample in the external circuit (*i.e.* the electron tunneling from the tip into the sample at the contact), which was ambiguous previously[5].

We determined the $\Delta V_s^*$ of the tribo-tunneling current using a null method. In a d.c. generator system like solar cell, the value of $V_{oc}$ can be determined by applying external bias to nullify the current output. **Figure 1**i-**1**l show the C-AFM current distribution with varying external bias $V_{bias}$ ($F$=500 nN). The $V_{oc}$ is estimated to be 300-400 mV when the current signal diminishes to ~0. Notably, the value of $\Delta V_s^*$ is 3~4 times higher than that of the $\Delta V_{CPD}$ under the same force (96 mV). The physical difference between $\Delta V_s^*$ and $\Delta V_{CPD}$ is more evident from the microscopic experimental data (**Figure 2**a), where a gold-coated spring connector with tip diameter ~59 μm is used to control the contact force in measurement. It can be seen that the sliding movement induce $V_{oc}$ of magnitude around 300~400 mV that immediately decays to zero under static condition (**Figure 2**b).

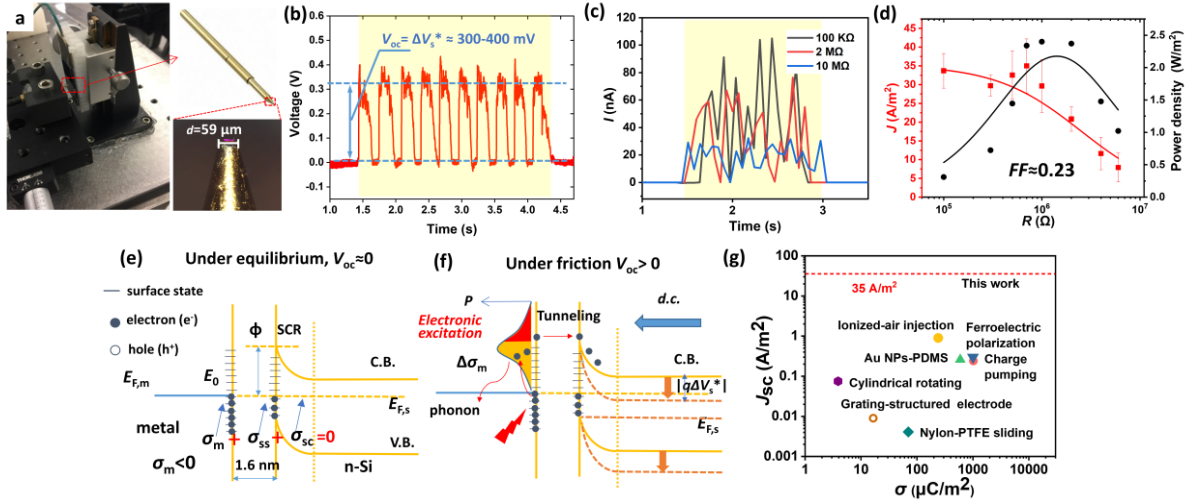

**Figure 2.** (a) Macroscopic measurement setup. (b) $V_{oc}$ output in the macroscopic measurement. The time periods of frictional movements are marked as yellow. The contact force $F$= 0.5 N (c) Current signal $I$ at electrical load R=100 kΩ, 1 MΩ and 10 MΩ. (d) Current density $J$ and power density as a function of $R$. The data are fitted with Eq. S3 and Eq. S4 in the supporting information; Proposed energy band diagram of the contact system (e) under equilibrium, n-type Si, (f) under friction, n-type Si; (g) The comparison of charge density $\sigma$ and short-circuit $J_{sc}$ reported in the conventional TENGs and this work. Note: the $\sigma$ refers to the dynamic, non-equilibrium interfacial excitation charge density in the tribo-tunneling systems, and the electrostatic, equilibrium surface charge density in the conventional TENGs, respectively. The references are listed in the reference group [28] and [29].

The current density ($J$) and power density as a function of electric load $R$ are collected under $F$=0.5 N and shown in **Figure 2**c. The short-circuit current $I_{sc}$ is measured to be ~100 nA, which corresponds to $J_{sc}$ ~35 A/m² considering a circular contact interface with diameter of ~50 μm (**Figure 2**a). Here, the parameter fill factor ($FF$), defined as the ratio of the maximum power to the product of $V_{oc}$ and $I_{sc}$ in solar cell, is adopted to characterize the quality of power generation in the tribo-tunneling system[19]:

$$FF = P_{max}/(I_{sc}V_{oc}), \quad (3)$$

Accordingly, the calculated FF is ~0.23, and the lower value can be attributed to the large contact resistance of the point-plane interface. Previously, we have reported that the $J_{sc}$ of ~2.5 A/m$^2$ and the power density of ~0.3 W/m$^2$ in the MIS system by multimeter probe with a tip diameter ~1 mm, and *FF* of ~0.34.[5] Therefore, it is suggested that the reducing tip radius could enhance the d.c. density, which is in good agreement with our AFM results of the tribo-tunneling systems.[4, 5] However, the internal energy conversion efficiency may be influenced by the increased contact resistance in the sharp point contact.

We shall now discuss the mechanism of the $\Delta V_s^*$ generation in sliding MIS contacts. Neglecting the bulk defects within the thin oxide layer, from charge neutrality we can express the interfacial charge as[16, 20]:

$$\sigma_m + \sigma_{ss} + \sigma_{sc} = 0, \quad (4)$$

where $\sigma_m$, $\sigma_{ss}$, and $\sigma_{sc}$ denotes the charge density at metal surface, Si-SiO$_x$ interface and Si surface depletion layer (SCR), respectively (**Figure 2**e). Empirically, the sign of $\sigma_m$ depends on the relative potential values of the contact materials in the triboelectric series, which is described by different contact electrification theories such as electron transfer, ion transfer, and mass transfer.[9] It is known from the KPFM results that $\sigma_m<0$, which corresponds to electron transfer from oxide surface to metal, leaving excess electrons on the metal side and holes on the oxide surface. This is understandable when the metal Fermi-level $E_F$ is lower than the $E_o$ of the oxide surface before contact.

Under frictional motion, atomic distance between the two contact materials may be "squeezed" periodically, considering the existence of microscopic nanogaps from surface roughness. At atomic level, periodically changing atomic distance is also expected by "stick-slip" Prandtl-Tomlinson model.[1, 21] Consequently, electronic excitation of energetic electrons/holes to higher/lower energy distribution spectrum may be induced, when the symmetry of interfacial electrostatic equilibrium is interrupted on and off.[22] Here we assume a non-equilibrium charge density increase $\Delta\sigma$ that can dissipate through (**Figure 2**f):

$$\Delta\sigma_{\text{m}} = \Delta\sigma_{\text{m,tunnel}} + \Delta\sigma_{\text{m,acc}} + \Delta\sigma_{\text{m,phonon}}, \quad (5)$$

where $\Delta\sigma_{\text{m,tunnel}}$, $\Delta\sigma_{\text{m,acc}}$, and $\Delta\sigma_{\text{m,phonon}}$ denotes the portion of $\Delta\sigma_{\text{m}}$ dissipating into tunneling current, surface charge accumulation, and heat, respectively. Among the three terms, and $\Delta\sigma_{\text{m,tunnel}}$ and $\Delta\sigma_{\text{m,acc}}$, are the results of non-adiabatic energy conversion into energetic carriers deviated from the Oppenheimer approximation.[22] For the electrons with sufficient energy to overcome the Schottky barrier (marked as red in the energy distribution spectrum in **Figure 2**f), they can pass through the thin oxide layer via quantum mechanical tunneling into the semiconductor side ("forward" tunneling), which corresponds to $\Delta\sigma_{\text{m,tunnel}}$. The ones with insufficient energy for tunneling, but enough energy for contact electrification (it also has an energy barrier as predicted by density function theory (DFT) calculation[17]) will be trapped at the surface states ($\Delta\sigma_{\text{m,acc}}$), or "back" tunneling into the metal side[9, 23]. The tribo-tunneling generator uses $\Delta\sigma_{\text{m,tunnel}}$ as the d.c. source, of which the tunneling current $I$ can be calculated by[22, 24]:

$$I(d) = \alpha \exp(-\beta d) \exp\left[-\frac{\Phi}{k_{\text{B}}T}\right], \quad (6)$$

where, $d$ is the oxide thickness, $\Phi$ is the Schottky energy barrier height, $k_{\text{B}}$ is the Boltzmann constant, and $T$ is the temperature. $\alpha$ and $\beta$ are parameters which are associated with surface state density and effective contact area. In Eq. 6, the first and second exponential component corresponds to the exponential decayed probability of quantum tunneling as a function of forbidden region distance (one-dimensional)[5] and the probability of thermionic emission of hot carrier through the Schottky barrier, respectively. A similar situation can be found in catalytic reaction of molecules on the Schottky junctions.[22, 25] The chemically induced electronic excitation is manifested as exoelectron emission, surface chemiluminescence, and "hot electron" tunneling current[22]—in analogy with tribo-exoelectron[26], triboluminescence[27], and the tribo-tunneling current of MIS junction in our case.

As summarized in **Figure 2**g, the $J$ of tunneling d.c. is about 3-4 orders' higher than dielectric a.c. density in conventional TENGs.[28, 29] Moreover, the current output exhibits a sustained feature under continuous or a relatively fast reciprocating sliding fashion, while the dielectric a.c. manifests itself as transient output as a function of time.[3, 5] In conventional TENGs, the maximum electrostatic $\Delta\sigma_m$ achieved by external charge injection is ~1 mC/m$^2$, which approaches the dielectric air break down limitation.[28] The quantification of $\Delta\sigma_m$ in the tribo-tunneling transport may need further scrutiny into the space where the voltage is built up. If we assume the voltage is built up capacitively across the 1.6 nm surface oxide, the dynamic $\Delta\sigma_m$ is estimated as high as 6.5 mC/m$^2$, using $V_{oc}$=300 mV, $d$=1.6 nm, $\varepsilon$=3.9, $\varepsilon_0$=8.85×10$^{-12}$ F/m. A more plausible scenario is that the $V_{oc}$ is exerted on the Si depletion layer with the width on the order of 100 nm. Accordingly, $\Delta\sigma_m$ would be on the order of 0.1-1 mC/m$^2$.

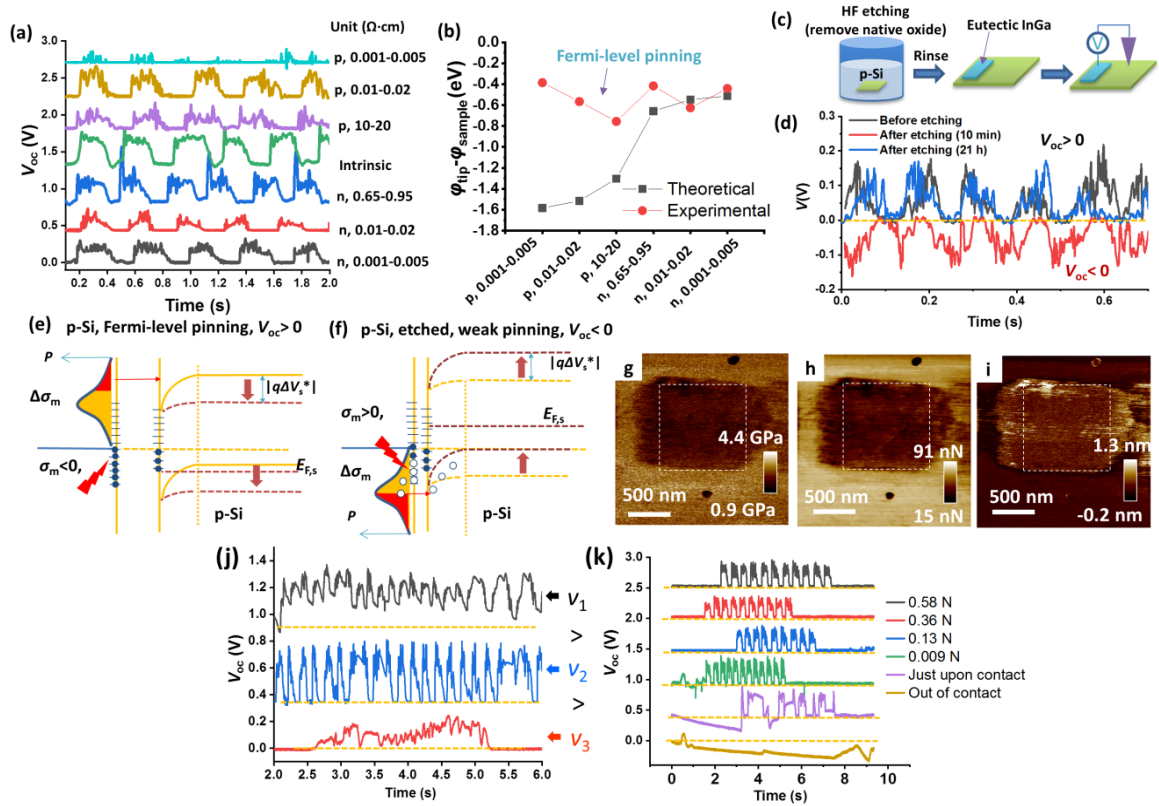

**Figure 3**. (a) Si doping type/concentration-dependent voltage output; The data sets are shifted in y-axis for better visualization. (b) Experimentally measured and theoretically predicted surface potential difference. (c) Surface oxide removal procedures via HF etching. (d) $V_{oc}$ signal

before, after (10 min and 21 h) etching; Proposed energy band diagram of the contact (c) under friction, p-type Si before etching, and (d) under friction, p-type Si after etching; Energy band diagram of the p- type Si contact system (e) before and (f) after HF etching. (g) DMT (Derjaguin-Muller-Toporov) modulus, (h) adhesion, (i) deformation distribution of the pre-rubbed area under $F$=200 nN. (j) Velocity-dependent voltage output. (k) Contact force-dependent voltage output; The data sets in (j) and (k) are shifted in y-axis for a better comparison. The zero voltage level for each set of data is marked as yellow dashed line.

The dependence of $V_{oc}$ signals on the Si doping type and resistivity are summarized in **Figure 3**a. It can be seen that the tip-surface electronic interaction exhibits only a weak dependency on Si doping type or concentration, which may be related to the Fermi-level pinning at the Si surface with 1-2 nm native oxide.[20, 30] The Fermi-level pinning effect can be inferred from the KPFM measurement of the surface potential of Si surfaces where the surface energy level $E_0$ remains relatively constant in all doping type/concentrations (**Figure 3**b). Specifically, the thin native oxide on the Si surface plays an important role in generating the interfacial electronic excitation, which is reflected in the increasing $V_{oc}$ output as function of increasing thickness of oxide deposited with atomic layer deposition (ALD).[5] For comparison we removed the native oxide on a 0.1-1 Ω·cm p-type Si by HF etch and measured the $V_{oc}$ signal (**Figure 3**c). The measured $V_{oc}$ direction reverses right after the etching (~5 min) and restores after 20 hrs of exposure to ambient air (oxide growth). This indicates that the electronic excitation induces opposite interfacial potential difference with and without surface native oxide (**Figure 3**d). It is assumed that the removal of surface oxide may weaken the Fermi-level pinning effect, which in turn reverse the surface potential difference between the metal and the p-Si surface resulting in the opposite $\sigma_m$ (>0, electron transfer from metal to sample surface) (**Figure 3**e-3f). As a result, $\Delta\sigma_m$ may be interpreted as increasing "hot" holes as illustrated in **Figure 3**f. This is in good agreement with the observed $V_{oc}$ direction reversal in the combination of contact materials for the tribo-tunneling systems with different work function.[5, 6]

Under Peakforce tapping mode, localized mechanical properties such as elastic modulus, deformation, and adhesion can be extracted from the force curve collected on each cycle (**Figure S**4 in Supporting Information). As shown in **Figure 3**g-3i, the reduced modulus and adhesion, and the increased deformation of the rubbed area indicates a possible Si-O surface bond rupture during the friction, increasing the surface state density and thus contributing to the $\Delta\sigma_m$. As proposed by Grunthaner, *et al*.[31], excited Si surface with native oxide may induce a large number of Si-O bond cleavages with stationary $Si^{3+}$ defects (holes) and mobile $O^-$ defects (electrons). These charge carriers may quickly recombine releasing heat or take part in electrons tunneling, which is in line with our above discussion from energy band diagram point of view.

The $V_{oc}$ outputs for three different sliding speeds $v$ ($v_1 > v_2 > v_3$) are shown on **Figure 3**j. It can be seen that the $V_{oc}$ under the slowest sliding shows a smaller value compared to the other two cases. When the sliding speed/frequency is fast enough (the case with $v_1$), the $V_{oc}$ output becomes continuous, which is in line with the reported result under a circular sliding mode.[5] Interestingly, the contact force dependent $V_{oc}$ measurements (**Figure 3**k) show that the rising potential difference is not affected by the contact force in a wide force range. Since the surface oxide is more rigid than organic materials, it may be less prone to a significant contact area change, which has been considered as a main reason why the apparent charge density of polymer material-based contacts is always enhanced with increasing force.[9] It is also noted that the $V_{oc}$ signal when the two materials are separated is fluctuated (**Figure 3**k). This resembles the $V_{oc}$ output of a conventional vertical mode TENG, where the voltage is now a result of work function difference as well as trapped triboelectric surface charges.[3]

In summary, it has been revealed that the tribo-tunneling transport in the metal-insulator-semiconductor sliding system is associated with the friction-induced electronic excitation at interface. The driving force of the tribo-tunneling transport and dielectric displacement current transport in conventional TNEGs is very different: conventional TENGs use electrostatically trapped surface charges as the voltage source, whereas the tribo-tunneling stems from the rising interfacial charge $\Delta\sigma_m$ under non-equilibrium condition, which is subsequently dissipated into tunneling current, surface trapped charges, or heat, depending on the energy of the excited electrons/holes. It has been demonstrated that the strong electronic excitation in the MIS sliding system can generate high d.c. current density ($J$~35 A/m$^2$), which is 3-4 orders' higher than the dielectric a.c. output in conventional TENGs. These basic understandings provided a new direction for optimizing and developing next-generation mechanical energy conversion technologies based on MIS sliding systems.

**Supporting Information**
Supporting Information is available from the Wiley Online Library or from the author.

**Acknowledgements**

This work was supported by the Canada Excellence Research Chair (CERC) program at the University of Alberta. The authors would like to acknowledge the support from Alberta Innovates-Technology Futures (AITF) Graduate Scholarship.

Received: ((will be filled in by the editorial staff))
Revised: ((will be filled in by the editorial staff))
Published online: ((will be filled in by the editorial staff))